\documentclass{ws-procs9x6}

\usepackage{epsfig,multicol,bbm}

\def\Journal#1#2#3#4{{\it #1} {\bf #2}, #3 (#4)}

\def\NP{{Nucl.\ Phys.}}
\def\PL{{Phys.\ Lett.}}
\def\PRL{Phys.\ Rev.\ Lett.}

\def\JHEP{{J.\ High Energy Phys.}}

\def\APJS{{Astrophys.\ J.\ Suppl.}}

\def\be{\begin{equation}}
\def\ee{\end{equation}}
\def\bea{\begin{eqnarray}}
\def\eea{\end{eqnarray}}
\newcommand{\befig}{\begin{figure}}
\newcommand{\efig}{\end{figure}}
\newcommand{\gtsim}{\;\lower-0.45ex\hbox{$>$}\kern-0.77em\lower0.55ex\hbox{$\sim$}\;}
\newcommand{\ltsim}{\;\lower-0.45ex\hbox{$<$}\kern-0.77em\lower0.55ex\hbox{$\sim$}\;}
\newcommand{\gl}{\ensuremath{\tilde{g}}}
\newcommand{\ax}{\ensuremath{\tilde{a}}}
\newcommand{\sq}{\ensuremath{\tilde{q}}}

\newcommand{\keV}{\mbox{keV}}
\newcommand{\MeV}{\mbox{MeV}}
\newcommand{\GeV}{\mbox{GeV}}

\begin{document}

\title{Thermal Production of Axinos\\ in the Early Universe}

\author{A.~BRANDENBURG and F.~D.~STEFFEN
\footnote{\uppercase{T}alk given by \uppercase{F.D.S.}\ 
at \uppercase{SEWM} 2004, 
\uppercase{H}elsinki, \uppercase{F}inland, \uppercase{J}une 16 -- 19, 2004.}
}

\address{DESY Theory Group,
Notkestrasse 85,
D-22603 Hamburg, Germany\\ 
E-mail: Arnd.Brandenburg@desy.de, Frank.Daniel.Steffen@desy.de}

\maketitle

\abstracts{We compute the thermal axino production rate in
  supersymmetric QCD to leading order in the gauge coupling. Using
  hard thermal loop resummation and the Braaten--Yuan prescription, we
  obtain a finite result in a gauge-invariant way, which takes into
  account Debye screening in the hot quark--gluon--squark--gluino
  plasma.  The relic axino density from thermal reactions in the early
  Universe is evaluated assuming the axino is the lightest
  supersymmetric particle and stable due to R-parity conservation.
  From the comparison with the WMAP results, we find that axinos could
  provide the dominant part of cold dark matter, for example, for an
  axino mass of 100~keV and a reheating temperature of $10^6\,\GeV$.}

\section{Introduction}
\label{Sec:Introduction}

In supersymmetric extensions of the standard model in which the strong
CP~problem is solved by the Peccei--Quinn (PQ) mechanism, the
axino~$\ax$ arises naturally as the fermionic superpartner of the
axion.\cite{Nilles:1981py+X} Thus, the axino is electrically and color
neutral and its interactions with the MSSM particles are suppressed by
the PQ~scale $f_a/N \gtsim 5 \times 10^9\,\GeV$. If the axino is the
lightest supersymmetric particle (LSP) and if $R$-parity is conserved,
axinos are stable and could provide the dominant part of cold dark
matter.

In this talk we present the computation of the thermal axino
production rate at high temperatures and discuss its cosmological
implications. With inflation governing the earliest moments of the
Universe, any initial population of axinos was diluted away and the
thermal production of axinos set in at the reheating
temperature~$T_R$. We restrict our investigation to $f_a/N > T_R
\gtsim 10^4\,\GeV$, where the U(1)$_{\mathrm{PQ}}$ symmetry is broken
and axino production from decays of particles out of equilibrium is
negligible.\cite{Covi:2001nw} The results presented are extracted from
Ref.~3 where more details can be found.

\section{Thermal Axino Production in Supersymmetric QCD}
\label{Sec:Axino_Production}

We concentrate on the axino--gluino--gluon interactions given by the
dimension-5 interaction term
\be
        {\mathcal L}_{\ax\gl g}= 
        i\,\frac{g^2}{64\pi^2 (f_a/N)}\,\bar{\ax}\,\gamma_5
        \left[\gamma^{\mu},\gamma^{\nu}\right]\,\gl^a\, G^a_{\mu\nu}
\label{Eq:L_agG}
\ee
with the strong gauge coupling $g$, the gluon field strength tensor
$G^a_{\mu\nu}$, the gluino $\gl$, and $N$ being the number of quarks
with PQ~charge. The resulting $2\to 2$ scattering processes in
supersymmetric QCD and the corresponding squared matrix elements are
listed in Table~\ref{Tab:diffcs}, where $s=(P_1 + P_2)^2$ and $t=(P_1
- P_3)^2$ with $P_1$, $P_2$, $P_3$, and $P$ referring to the particles
in the given order. Working in the limit, $T \gg m_i$, the masses of
all particles involved have been neglected.  Sums over initial and
final spins have been performed. For quarks and squarks, the
contribution of a single chirality is given.
\begin{table}[b!]\renewcommand{\arraystretch}{1.2}
\tbl{Squared matrix elements for axino ($\ax$) production in two-body processes involving quarks ($q_i$), squarks ($\sq_i$), gluons ($g^a$), and gluinos ($\gl^a$) in the high-temperature limit, $T \gg m_i$, with the SU(3) color matrices $f^{abc}$ and $T^a_{ji}$}
{\footnotesize
\begin{tabular}{@{}rcl@{}}
\hline\\[-2.5ex]
{} & \quad process $i$ \quad & 
\quad\quad $|{\mathcal M}_i|^2/\frac{g^6}{128\pi^4(f_a/N)^2}$ \quad\quad \\[1.5ex]
\hline
\\[-2.5ex]
A & \quad $g^a + g^b \rightarrow \gl^c + \ax $ \quad & 
\quad $4(s+2t+2\frac{t^2}{s})|f^{abc}|^2$ 
\\
B & \quad $g^a + \gl^b \rightarrow g^c + \ax $ \quad & 
\quad $-4(t+2s+2\frac{s^2}{t})|f^{abc}|^2$
\\
C & \quad $\sq_i  + g^a \rightarrow q_j + \ax $ \quad & 
\quad $2s|T^a_{ji}|^2$
\\
D & \quad $g^a + q_i \rightarrow \sq_j + \ax $ \quad & 
\quad $-2t|T^a_{ji}|^2$
\\
E & \quad $\bar{\sq}_i + q_j \rightarrow g^a + \ax $ \quad & 
\quad $-2t|T^a_{ji}|^2$
\\
F & \quad $\gl^a + \gl^b \rightarrow \gl^c + \ax $ \quad & 
\quad $-8\frac{(s^2+st+t^2)^2}{st(s+t)}|f^{abc}|^2$
\\
G & \quad $q_i + \gl^a \rightarrow q_j + \ax $ \quad & 
\quad $-4(s+\frac{s^2}{t})|T^a_{ji}|^2$
\\
H & \quad $\sq_i + \gl^a \rightarrow \sq_j + \ax $ \quad & 
\quad $-2(\frac{t}{2}+2s+2\frac{s^2}{t})|T^a_{ji}|^2$
\\
I & \quad $q_i + \bar{q}_j \rightarrow \gl^a + \ax $ \quad & 
\quad $-4(t+\frac{t^2}{s})|T^a_{ji}|^2$
\\
J & \quad $\sq_i + \bar{\sq}_j \rightarrow \gl^a + \ax $ \quad & 
\quad $2(\frac{s}{2}+2t+2\frac{t^2}{s})|T^a_{ji}|^2$
\\[1.0ex] 
\hline
\end{tabular}\label{Tab:diffcs} }
\vspace*{-13pt}
\end{table}

The processes B, F, G, and H lead to a logarithmic collinear
singularity due to $t$-channel and $u$-channel exchange of soft
(massless) gluons. Here screening effects in the hot
quark--gluon--squark--gluino plasma (QGSGP) have to be taken into
account. With hard thermal loop (HTL)
resummation,\cite{Braaten:1989mz} this can be done systematically in a
gauge-invariant way. Following the Braaten--Yuan
prescription,\cite{Braaten:1991dd} we introduce a momentum scale
$k_{\rm cut}$ such that $gT \ll k_{\rm cut} \ll T$ in the weak
coupling limit, $g \ll 1$. This separates soft gluons with momentum
transfer of order $gT$ from hard gluons with momentum transfer of
order $T$. In the region of soft momentum transfer, $k<k_{\rm cut}$,
we use the HTL-resummed gluon propagator, which takes into account
Debye screening in QGSGP in terms of the supersymmetric thermal gluon
mass $m_g = g T \sqrt{(N_c+n_f)/6}$, where $N_c=3$ and $n_f=6$ are
respectively the number of colors and color triplet and anti-triplet
chiral multiplets.  Computing the imaginary part of the thermal axino
self-energy with the ultraviolet (UV) momentum cutoff $k_{\rm cut}$,
the soft contribution to the production rate of axinos with energies
$E \gtsim T$ is obtained
\be
\label{softres}
        \left.\frac{d\Gamma^{\tilde{a}}}{d^3p}\right|_{\rm soft}
        = f_F(E)\frac{3g^4(N_c^2-1)m_g^2T}{4096\pi^8(f_a/N)^2}
        \left[
        \ln\left(\frac{k_{\rm cut}^2}{m_g^2}\right)-1.379
        \right],
\ee
where $f_F(E)=[\exp(E/T+1)]^{-1}$. In the region of hard momentum
transfer, $k>k_{\rm cut}$, the bare gluon propagator can be used. From
the squared matrix elements of the $2 \to 2$ processes (cf.\ 
Table~\ref{Tab:diffcs}) weighted with appropriate multiplicities,
statistical factors, and phase space densities, the corresponding hard
contribution can then be obtained conveniently
\be
\label{hardres} 
\left.\frac{d\Gamma_{\ax}}{d^3p}\right|_{\rm hard}
= 
\frac{g^6\,(N_c^2-1)}{32\,\pi^4\,(f_a/N)^2}
\left[ (N_c+n_f)\frac{T^3\, f_F(E)}{128\pi^4}\, 
\ln\bigg(\frac{2^{1/3}T}{k_{\rm cut}}\bigg)
+ ...
\right]
\ee
with $k_{\rm cut}$ as the infrared (IR) cutoff for gluon momentum
transfers. The expression represented by the ellipses can be found in
Ref.~3. It is independent of $k_{\rm cut}$. Thus, by summing the soft
and hard contributions, the artificial $k_{\rm cut}$ dependence
cancels and the finite leading order rate is obtained
%
\bea
\label{Eq:total_rate_axino}
\frac{d\Gamma_{\ax}}{d^3p}=
\left.\frac{d\Gamma_{\ax}}{d^3p}\right|_{\rm soft}
+\left.\frac{d\Gamma_{\ax}}{d^3p}\right|_{\rm hard}.
\eea
In Fig.~\ref{Fig:Axino_Production}a, we show the normalized thermal
axino production rate $1/\Gamma_{\ax}\,d\Gamma_{\ax}/d(E/T)$ as a
function of $E/T$ for temperatures of $T=10^6\,\GeV$, $10^7\,\GeV$,
$10^8\,\GeV$, and $10^9\,\GeV$. For $E\ltsim T$, we expect
modifications as our computation is restricted to the thermal
production of hard ($E \gtsim T$) axinos.\cite{Brandenburg:2004du}
Here and below the 1-loop running of the strong coupling in the MSSM
is taken into account by replacing $g$ with
$g(T)=[g^{-2}(M_Z)\!+\!3\ln(T/M_Z)/(8\pi^2)]^{-1/2}$, where the value
of the strong coupling at the Z-boson mass $M_Z$,
$g^2(M_Z)/(4\pi)=0.118$, is used as input. For $E/T\lesssim 2$, the
rate turns negative except for very high temperatures. With
$g(T\!=\!10^6\,\GeV)=0.986$ and $g(T\!=\!10^9\,\GeV)=0.880$, this
unphysical behavior follows from the extrapolation of our result from
$g\ll 1$ to $g\approx 1$. As higher-order corrections might become
sizable for $T \ltsim 10^6\,\GeV$, new techniques are needed to
compute the thermal production rate reliably also for $g\gtsim 1$.  At
present, however, our result derived in a gauge-invariant way
supersedes previous gauge-dependent and cutoff-dependent
estimates.\cite{Covi:2001nw}
\befig[t] 
\epsfig{figure=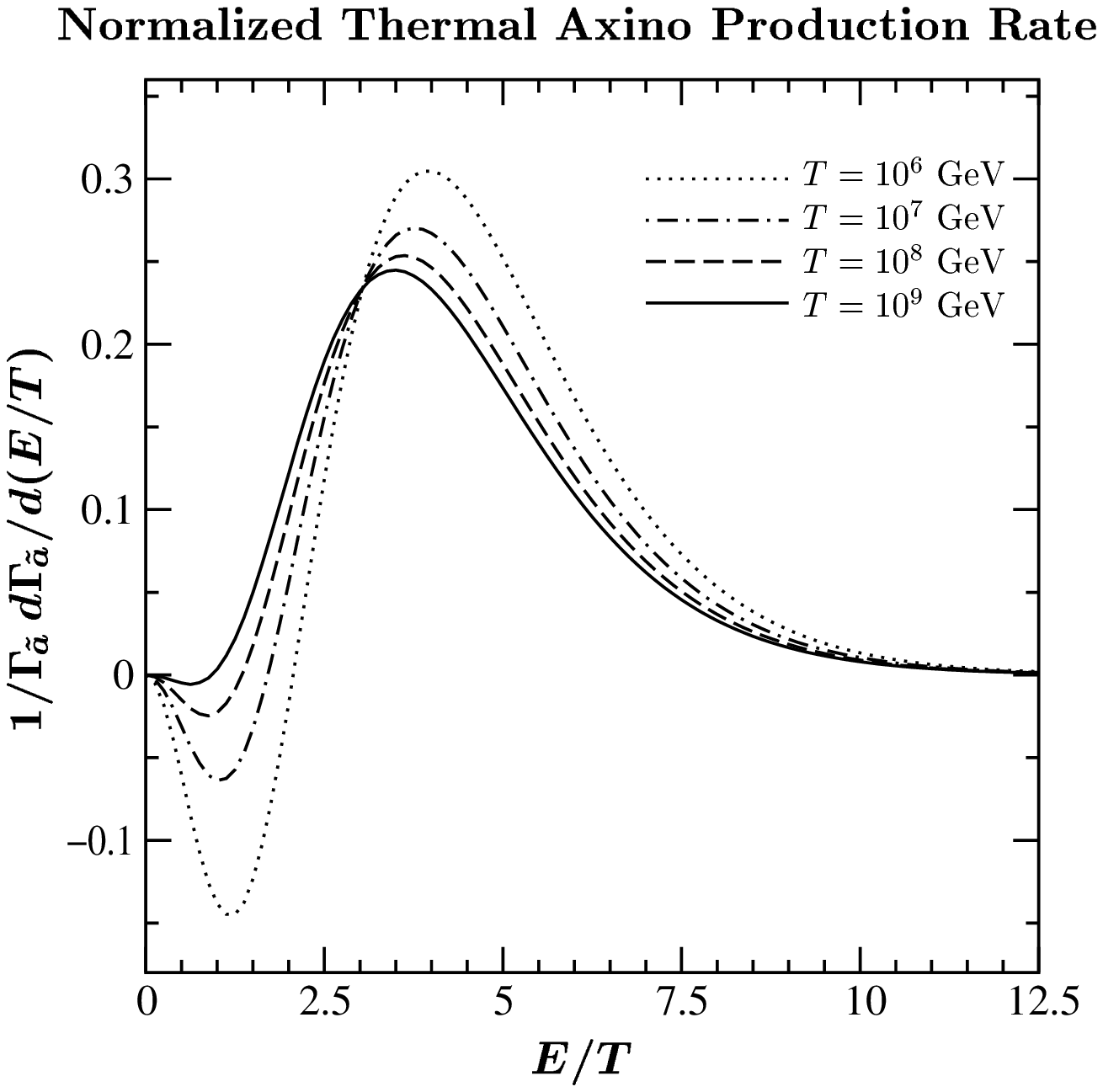,width=5.7cm}
\hfill
\epsfig{figure=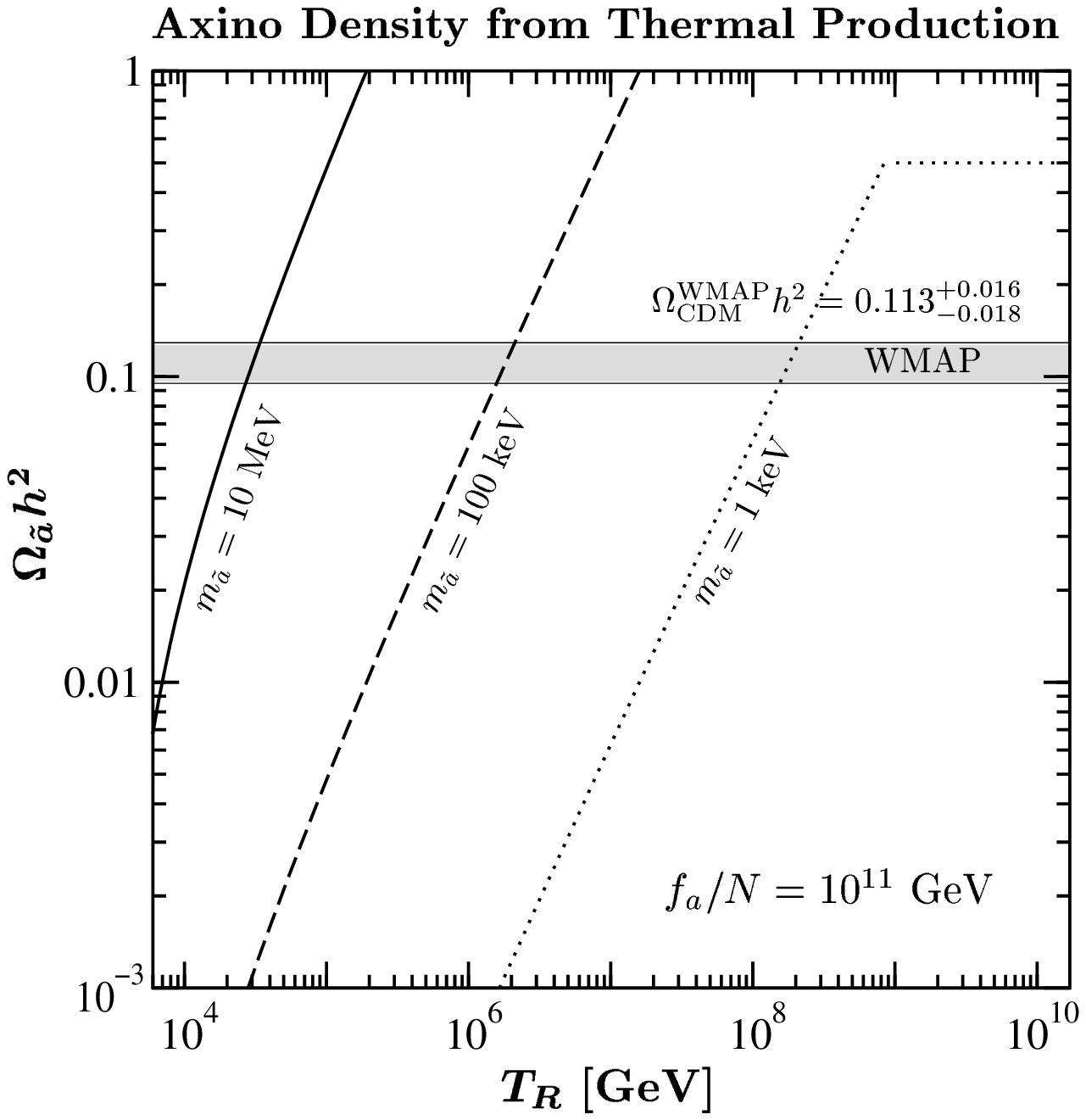,width=5.5cm}
\caption{
  (a) The normalized thermal axino production rate
  $1/\Gamma_{\ax}\,d\Gamma_{\ax}/d(E/T)$ as a function of $E/T$ for
  $T=10^6\,\GeV$ (dotted line), $10^7\,\GeV$ (dash-dotted line),
  $10^8\,\GeV$ (dashed line), and $10^9\,\GeV$ (solid line). The
  curves result from~(\ref{Eq:total_rate_axino}) derived for $E \gtsim
  T$. (b) The axino density parameter $\Omega_{\ax}h^2$ as a function
  of $T_R$ for $f_a/N = 10^{11}\,\GeV$ and $m_{\ax} = 1~\keV$ (dotted
  line), $100~\keV$ (dashed line), and $10~\MeV$ (solid line). The
  grey band indicates the WMAP result on the relic density of cold
  dark matter (2$\sigma$ error).}
\label{Fig:Axino_Production}
\efig
%
\section{Stable Axinos as Dark Matter in the Universe}
\label{Sec:Axino_Dark_Matter}

By comparing the axino interaction rate for $f_a/N = 10^{11}\,\GeV$
with the Hubble parameter $H$ for the early radiation-dominated epoch,
an axino decoupling temperature of $T_D\approx 10^9\,\GeV$ is
estimated.\cite{Rajagopal:1990yx} For $T_R < T_D$, axinos have not
been in thermal equilibrium after inflation. Here the evolution of the
axino number density $n_{\ax}$ with cosmic time $t$ is described by
the Boltzmann equation with a collision term $C_{\ax}$ accounting for
both the axino production and disappearance processes in the
primordial plasma,
\be
    \frac{dn_{\ax}}{dt} + 3 H n_{\ax} = C_{\ax}
    \ .
\label{Eq:Boltzmann}
\ee
The disappearance processes can be neglected for $T_R$ sufficiently
below $T_D$. Then, by integrating the production
rate~(\ref{Eq:total_rate_axino}), we obtain the collision term
\be
        C_{\ax}
        \approx
        \frac{(N_c^2 - 1)}{(f_a/N)^2} 
        \frac{3\zeta(3) g^6 T^6}{4096\pi^7}
        \Bigg[\ln\left(\frac{1.380\,T^2}{m_g^2}\right)(N_c+n_f) 
        + 0.4336\,n_f \Bigg]
        \ ,
\label{Eq:C_axino}
\ee
where $\zeta(3)\approx 1.2021$. Assuming conservation of entropy per
comoving volume, the Boltzmann equation can be solved analytically.
The resulting present ($t_0$) axino density parameter $\Omega_{\ax}h^2
= m_{\ax} n_{\ax}(t_0)h^2/\rho_c$ with $\rho_c/h^2 = 3.6\times
10^{-9}\,\GeV$ depends on the axino mass $m_{\ax}$, the PQ scale
$f_a/N$, and the reheating temperature $T_R$ in the following way
\be
        \Omega_{\ax}h^2
        = 5.5\,g^6 \ln\left( \frac{1.108}{g}\right) 
        \bigg(\frac{m_{\ax}}{0.1~\GeV}\bigg)\!
        \left(\frac{10^{11}\,\GeV}{f_a/N}\right)^{\! 2}\!\!
        \left(\frac{T_R}{10^4\,\GeV}\right) \, ,
\label{Eq:Omegah2_axino}
\ee
where $g = g(T_R)$. In Fig.~\ref{Fig:Axino_Production}b, this result
is illustrated as a function of $T_R$ for $f_a/N = 10^{11}\,\GeV$ and
$m_{\ax} = 1~\keV$, $100~\keV$, and $10~\MeV$.  For $T_R$ above
$T_D$, axinos are in thermal equilibrium so that $\Omega_{\ax}h^2$ is
independent of $T_R$ as shown for $m_{\ax} = 1~\keV$ by the
horizontal line. There will be a smooth transition instead of a kink
once the axino disappearance processes are taken into account.  The
grey band indicates the WMAP result\cite{Spergel:2003cb} on the cold
dark matter density (2$\sigma$ error)
$\Omega_{\mathrm{CDM}}^{\mathrm{WMAP}} h^2 = 0.113^{+0.016}_{-0.018}$.

\section{Conclusion}
\label{Sec:Conclusion}

For $m_{\ax}=100~\keV$ and $T_R \approx 10^6\,\GeV$, the relic axino
density (obtained with $f_a/N = 10^{11}\,\GeV$) agrees with the WMAP
result on the cold dark matter density.  Although relatively light for
being cold dark matter, axinos with $m_{\ax}=100~\keV$ could still
explain large-scale-structure formation, the corresponding power
spectrum, and the early reionization observed by WMAP. Thus, as far as
abundance and structure formation are concerned, axinos from thermal
production in the early Universe are indeed a viable solution of the
cold dark matter problem.\cite{Brandenburg:2004xh} This conclusion has
already been drawn in Ref.~2 based on a gauge-dependent and
cutoff-dependent estimate of the relic axino abundance from thermal
production. Since our result for $\Omega_{\ax}h^2$ is smaller by a
factor of ten, the upper limit on $T_R$ for a given value of $m_{\ax}$
is relaxed by one order of magnitude. This can be important for models
of inflation and for the understanding of the baryon asymmetry in the
Universe.  Already $T_R \approx 10^6\,\GeV$ is relatively small and
excludes, for example, thermal leptogenesis.

%
%
%
%


\begin{thebibliography}{0}

%
\bibitem{Nilles:1981py+X}
%
H.~P.~Nilles and S.~Raby,
\Journal{\NP}{B198}{102}{1982};\\
%
J.~E.~Kim and H.~P.~Nilles,
\Journal{\PL}{B138}{150}{1984}.

%
\bibitem{Covi:2001nw}
L.~Covi {\it et al.},
\Journal{\JHEP}{05}{033}{2001}.

%
\bibitem{Brandenburg:2004du}
A.~Brandenburg and F.~D.~Steffen,
arXiv: hep-ph/0405158.

%
\bibitem{Braaten:1989mz}
E.~Braaten and R.~D.~Pisarski,
\Journal{\NP}{B337}{569}{1990}.
%
\bibitem{Braaten:1991dd}
E.~Braaten and T.~C.~Yuan,
\Journal{\PRL}{66}{2183}{1991}.

%
\bibitem{Rajagopal:1990yx}
K.~Rajagopal, M.~S.~Turner, and F.~Wilczek,
\Journal{\NP}{B358}{447}{1991}.

%
\bibitem{Spergel:2003cb}
D.~N.~Spergel {\it et al.},
\Journal{\APJS}{148}{175}{2003}.

%
\bibitem{Brandenburg:2004xh}
A.~Brandenburg and F.~D.~Steffen,
arXiv:hep-ph/0406021.

\end{thebibliography}
\end{document}